\documentclass[prl,twocolumn,showpacs,superscriptaddress,floatfix, reprint]{revtex4}

\usepackage[OT1]{fontenc}
\usepackage[usenames,dvipsnames]{color}
\usepackage[latin1]{inputenc}
\usepackage[english]{babel}
\usepackage{graphicx}
\usepackage{color}
\usepackage[Gray,squaren]{SIunits}
\usepackage{xspace}
\usepackage{upgreek}
\usepackage{ulem}
\usepackage{epstopdf}
\usepackage{amssymb,amsmath,verbatim,ulem}

\newcommand{\ket}[1]{|{#1}\rangle}
\newcommand{\bra}[1]{\langle{#1}|}

\begin{document}

\title{Highly sensitive atomic based MW interferometry }
\author{Dangka Shylla, Elijah Ogaro, and Kanhaiya Pandey}
\affiliation{Department of Physics, Indian Institute of Technology Guwahati, Guwahati, Assam 781039, India }
 \email{kanhaiyapandey@iitg.ernet.in}


%\affiliation{
%\affiliation{
%\affiliation{%


\date{\today{}}

\begin{abstract}
We theoretically study a scheme to develop an atomic based MW interferometry using the Rydberg states in Rb. Unlike the traditional MW interferometry, this scheme is not based upon the electrical circuits,  hence the sensitivity of the phase and the amplitude/strength of the MW field is not limited by the Nyquist thermal noise. Further this system has great advantage due to its very high bandwidth, ranging from radio frequency (RF), micro wave (MW) to terahertz regime. In addition, this is \textbf{orders of magnitude} more sensitive to field strength as compared to the prior demonstrations on the MW electrometry using the Rydberg atomic states. However previously studied atomic systems are only sensitive to the field strength but not to the phase and hence this scheme provides a great opportunity to characterize the MW completely including the propagation direction and the wavefront. This study opens up a new dimension in the Radar technology such as in synthetic aperture radar interferometry. The MW interferometry is based upon a six-level loopy ladder system involving the Rydberg states in which two sub-systems interfere constructively or destructively depending upon the phase between the MW electric fields closing the loop. 
\end{abstract}

%\pacs{42.50.Md, 42.25.Dd}

\maketitle

%\tableofcontents

Atomic based standards such as time and length is already adopted and established due to their high reproducibility, accuracy, resolution and stability \cite{HAL06}.
Atoms have also been successfully used for DC and AC (MW and RF) magnetometry, reaching impressive sensitivity and spatial resolutions \cite{BUR07,LRD13,HDM13,HGP15}. Inspired by these successes recently, the atom based MW and RF electrometry has been investigated using the Rydberg states of the atoms \cite{KSB10,SSK12, SSK13,HYL16,KFK17,JHH17}. The success of these experiments for high sensitive electrometry is due to property of the Rydberg states i.e. availability of closely spaced levels (in the range of MW and RF region) with very high electric polarizability.
The electrometry is based upon the phenomenon of electromagnetically induced transparency (EIT) in which the absorption property of a probe laser is altered in the presence of control lasers and MW (or RF) field  in a four level system. EIT is sensitive to the field's strength, frequency and the polarization and so the electrometry.
%But EIT is not sensitive to the absolute phase phase stability of the lasers and control MW field, but not to the absolute phase.  

An oscillating electro-magnetic field i.e. MW electric field is characterized by it's strength/amplitude, frequency, polarization and the phase. Phase of the MW fields is detected using traditional MW interferometry which is based upon the electrical circuit, whose performance is greatly limited by its bandwidth and the Nyquist thermal noise \cite{KIY81,ITW98,IVT09}. Further the previously studied atomic based MW electrometry is not phase sensitive as EIT in a simple multilevel system, happens to be insensitive to the absolute phase of probe and the control fields but it's robustness depends upon the phase stability \cite{HSB08}.

% The phase of the EM field is generally inferred through the interference effect by another EM field having same frequency.   
%Recently atom has been used to detect the microwave (MW) and the RF field's amplitude, frequency and the polarization with very good stability and reproducibility \cite{SSK13, JSA12, JYR16} due to long lifetime and large polarizability of the Rydberg states. This is based upon the electromagnetically induced transparency (EIT) in a four level Rb atom in which the absorption of the probe laser is modified in the presence of the MW or RF field in a four level system. Again in this scheme the absorption of the probe laser does not depend upon the absolute phase of the any of the control field i.e. the optical control laser as well the MW or RF control field.  

Here, we explore a six-level loopy ladder system in which the absorption property of the probe laser has phase dependency on the MW fields forming a closed loop. This proposed system is also an order of magnitude more sensitive to field strength in comparison to the previously explored system. This is based upon the interference between two sub-systems driven by the MW fields forming the loop.

\begin{figure}
   \begin{center}
      \includegraphics[width =8cm]{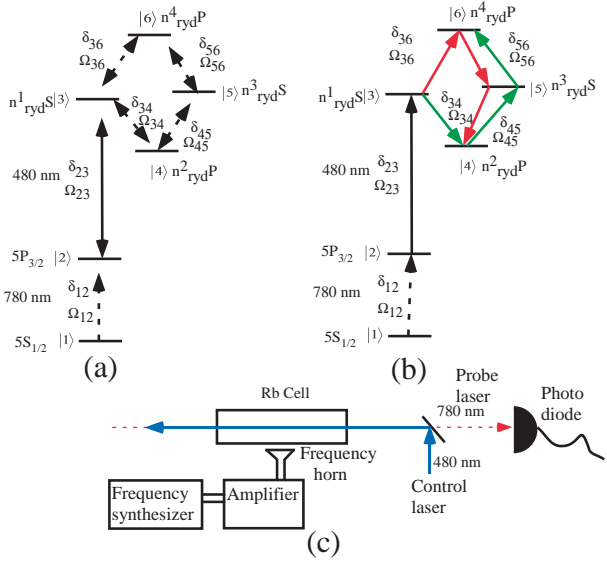}
      \caption{(Color online). (a) The energy level diagram for  loopy ladder system. (b) Transitions shown by the red and green arrow lines are the two sub-system to close the loop. The probe laser (dotted black arrow line) and the control laser (solid black arrow line) are part of both the sub-system. (c) The typical experimental set up for the phase dependent MW or RF electrometry.}
      \label{CLladdersys}
   \end{center}
\end{figure}

The considered six-level system is shown in Fig. \ref{CLladdersys}a.
In this system the electric field, associated with the transition $\ket{i}$ $\rightarrow$ $\ket{j}$ is $E_{ij}e^{i(\omega_{ij}t+\phi_{ij})}$, where $E_{ij}$ is amplitude, $\omega_{ij}$ is the frequency and $\phi_{ij}$ is the phase. We define
Rabi frequency $\Omega_{ij} =d_{ij}E_{ij}e^{i\phi_{ij}}/\hbar$ for the transition $\ket{i}$ $\rightarrow$ $\ket{j}$ having the dipole moment matrix element $d_{ij}$. The Rabi frequencies of the probe and the control lasers are $\Omega_{12}$ and $\Omega_{23}$ respectively, whereas $\Omega_{34}$, $\Omega_{45}$, $\Omega_{56}$ and $\Omega_{36}$ are the Rabi frequencies of the control MW fields. It is important to note here that the phase of $\Omega_{36}$ is to be characterized w.r.t to the reference MW fields $\Omega_{34}$, $\Omega_{45}$ and $\Omega_{56}$.    

The typical experimental setup for phase dependent MW electrometry is shown in Fig.\ref{CLladdersys}(c) in which a probe laser at 780~nm and a control laser at 480~nm are counter-propagating inside the Rb cell. The four MW control fields are generated by a single frequency synthesizer having arrangements of controlling the frequency, phase and the amplitude. The output of the frequency synthesizer is amplified and fed to MW horn. All four MW fields are propagating perpendicular to the probe and the control lasers with a uniform phase inside the Rb cell.

% Figure 1 %%%%%%%%%%%%%%%%%%%%%%%
%%%%%%%%%%%%%%%%%%%%%%%%%%%%%%%%%%

%The bare atomic Hamiltonian for above mentioned six-level system can be written as
%\begin{equation}
%H_0=\sum_{j=1}^{6}\hbar\omega_j\ket{j}\bra{j}
%\end{equation}
%The interaction Hamiltonian can be written as
The total Hamiltonian for this system is given as
\begin{align}
&H=\bigg[\sum^5_{i=1}\frac{\hbar\Omega_{i,i+1}}{2}\left(e^{i\omega_{i,i+1}t}+e^{-i\omega_{i,i+1}t}\right)\ket{i}\bra{i+1}\nonumber\\
&+\frac{\hbar\Omega_{36}}{2}\left(e^{i\omega_{36}t}+e^{-i\omega_{36}t}\right)\ket{3}\bra{6}+h.c.\Big]+\sum_{j=1}^{6}\hbar\omega_j\ket{j}\bra{j}\nonumber
\end{align}
%Hence, the total Hamiltonian will be
%\begin{equation}
%H=H_0+H_I
%\end{equation}
In the rotating frame with rotating wave approximation the above Hamiltonian will be
\begin{align}
H&=\hbar[0\ket{1}\bra{1}-\delta_{12}\ket{2}\bra{2}-(\delta_{12}+\delta_{23})\ket{3}\bra{3}\nonumber \\
&-(\delta_{12}+\delta_{23}-\delta_{34})\ket{4}\bra{4}-(\delta_{12}+\delta_{23}-\delta_{34}+\delta_{45})\ket{5}\bra{5}\nonumber \\
&-(\delta_{12}+\delta_{23}-\delta_{34}+\delta_{45}+\delta_{56})\ket{6}\bra{6}+\frac{\Omega_{12}}{2}\ket{1}\bra{2} \nonumber \\
&+\frac{\Omega_{23}}{2}\ket{2}\bra{3}+\frac{\Omega_{34}}{2}\ket{3}\bra{4}+\frac{\Omega_{45}}{2}\ket{4}\bra{5}+\frac{\Omega_{56}}{2}\ket{5}\bra{6} \nonumber \\
&+\frac{\Omega_{36}}{2}e^{i(\delta_{34}-\delta_{45}-\delta_{56}+\delta_{36})t}\ket{3}\bra{6}+h.c.\Big]\nonumber
\end{align}
where, $\hbar\omega_j$ is the energy of the state $\ket{j}$, $\delta_{12}=\omega^L_{12}-(\omega_2-\omega_1)$, $\delta_{23}=\omega^L_{23}-(\omega_3-\omega_2)$  are the detunings of the probe and control lasers, $\delta_{34}=\omega^L_{34}-(\omega_3-\omega_4)$, $\delta_{45}=\omega^L_{45}-(\omega_5-\omega_4)$, $\delta_{56}=\omega_{56}-(\omega_6-\omega_5)$ and $\delta_{36}=\omega_{36}-(\omega_6-\omega_3)$  are the detunings for the MW fields for the respective transitions.
The time evolution of the density matrix, $\rho$ is given as $\dot{\rho}=-\frac{i}{\hbar}[H, \rho]-\frac{1}{2}\{\Gamma,\rho\}$ where, $\Gamma$ is the relaxation matrix.\\
In general, the Hamilitonian $H$ is time dependent except for a particular condition when $\delta_{34}-\delta_{45}-\delta_{56}+\delta_{36}=0$.  
In the case of weak probe approximation there will be no population transfer and hence the time evolution of the population i.e. the diagonal terms of the density matrix such as $\rho_{11}$, $\rho_{22}$, $\rho_{33}$, $\rho_{44}$, $\rho_{55}$, and $\rho_{66}$  can be ignored with $\rho_{11}\approx1$,  $\rho_{22}\approx$ $\rho_{33}\approx$ $\rho_{44}\approx$ $\rho_{55}\approx$ $\rho_{66}\approx0$. The off-diagonal terms under this approximation $\rho_{ij}$=$\rho_{ji}\approx0$  for $i=2; j=3,4,5,6$ and $i=3; j=4,5,6$ and $i=4; j=5,6$ and $i=5; j=6$. 
In the case of steady state (i.e. $\dot{\rho_{ij}}=0$ for all the $i$ and $j$) and under weak probe approximation we get the following solution for $\rho_{12}$
\begin{align}
\rho_{12}&=\frac{\frac{i}{2}\frac{\Omega_{12}}{\gamma_{12}}}{1+\frac{\frac{1}{4}\frac{|\Omega_{23}|^2}{\gamma_{12}\gamma_{13}}}{1+\textrm{\textcolor{red}{EITATA1}}+\textrm{\textcolor{green}{EITATA2}}+\textrm{Interference}}}\\
\textrm{Where,} \nonumber \\
\textrm{\textcolor{green}{EITATA1}}&=\frac{\frac{1}{4}\frac{|\Omega_{34}|^2}{\gamma_{13}\gamma_{14}}}{1+\frac{\frac{1}{4}\frac{|\Omega_{45}|^2}{\gamma_{14}\gamma_{15}}}{1+\frac{1}{4}\frac{|\Omega_{56}|^2}{\gamma_{15}\gamma_{16}}}};\textrm{\textcolor{red}{EITATA2}}=\frac{\frac{1}{4}\frac{|\Omega_{36}|^2}{\gamma_{13}\gamma_{16}}}{1+\frac{\frac{1}{4}\frac{|\Omega_{56}|^2}{\gamma_{15}\gamma_{16}}}{1+\frac{1}{4}\frac{|\Omega_{45}|^2}{\gamma_{14}\gamma_{15}}}}\nonumber\\
\textrm{Interference}&=\frac{\frac{\frac{1}{16}\frac{|\Omega_{34}||\Omega_{45}||\Omega_{56}||\Omega_{36}|e^{i\phi}}{\gamma_{13}\gamma_{14}\gamma_{15}\gamma_{16}}}{1+\frac{1}{4}\frac{|\Omega_{56}|^2}{\gamma_{15}\gamma_{16}}}}{1+\frac{\frac{1}{4}\frac{|\Omega_{45}|^2}{\gamma_{14}\gamma_{15}}}{1+\frac{1}{4}\frac{|\Omega_{56}|^2}{\gamma_{15}\gamma_{16}}}}+\frac{\frac{\frac{1}{16}\frac{|\Omega_{34}||\Omega_{45}||\Omega_{56}||\Omega_{36}|e^{-i\phi}}{\gamma_{13}\gamma_{14}\gamma_{15}\gamma_{16}}}{1+\frac{1}{4}\frac{|\Omega_{45}|^2}{\gamma_{14}\gamma_{15}}}}{1+\frac{\frac{1}{4}\frac{|\Omega_{56}|^2}{\gamma_{15}\gamma_{16}}}{1+\frac{1}{4}\frac{|\Omega_{45}|^2}{\gamma_{14}\gamma_{15}}}}\nonumber
\label{ana1}
\end{align}
where, $\phi=\phi_{36}-\phi_{34}-\phi_{45}-\phi_{56}$; $\gamma_{12}=-\left[\gamma^{dec}_{12}+i\delta_{12}\right]$,\\
$\gamma_{13}=-\left[\gamma^{dec}_{13}+i\left(\delta_{12}+\delta_{23}\right)\right]$,
\\ $\gamma_{14}=-\left[\gamma^{dec}_{14}+i\left(\delta_{12}+\delta_{23}-\delta_{34}\right)\right]$, \\
$\gamma_{15}=-\left[\gamma^{dec}_{15}+i\left(\delta_{12}+\delta_{23}-\delta_{34}+\delta_{45}\right)\right]$,\\
$\gamma_{16}=-\left[\gamma^{dec}_{16}+i\left(\delta_{12}+\delta_{23}-\delta_{34}+\delta_{45}+\delta_{56}\right)\right]$.\\
$\gamma^{dec}_{ij}$ is decoherence rate between level $\ket{i}$ and $\ket{j}$. We take the value of $\gamma^{dec}_{12}=2\pi\times$3.05 MHz, which includes natural radiative decay of excited state, $\Gamma_2=2\pi\times$6MHz and the 780nm laser linewidth of $2\pi\times$50kHz. We also take $\gamma^{dec}_{13}=\gamma^{dec}_{14}=\gamma^{dec}_{15}=\gamma^{dec}_{16}=\gamma^{dec}=2\pi\times$100kHz mainly dominated by the laser linewidths of 780nm and the 480nm as compared to the radiative decay rate (=2$\pi$$\times$1kHz) of the Rydberg states $\ket{3}$, $\ket{4}$, $\ket{5}$ and $\ket{6}$ \cite{SSK13}. In order to check the stringency of the $\gamma^{dec}$ on the phase sensitivity, we set $\gamma^{dec}=2\pi\times$500kHz in another case.

In order to verify the approximation made above, we have checked the analytical solution of $\rho_{12}$ given by the Eq. [1] and the complete numerical solution in the steady state for various values of control fields and detunings. It has excellent agreement between complete numerical and approximated analytical solution.
The closed loop can be realized by two sub-systems $\ket{3}$$\rightarrow$$\ket{4}$$\rightarrow$$\ket{5}$$\rightarrow$$\ket{6}$ and $\ket{3}$$\rightarrow$$\ket{6}$$\rightarrow$$\ket{5}$$\rightarrow$$\ket{4}$ shown with red and green arrows respectively sharing a common $\ket{1}$$\rightarrow$$\ket{2}$$\rightarrow$$\ket{3}$ ladder system as shown in Fig.\ref{CLladdersys}b.
The control laser $\Omega_{23}$ causes reduction in the absorption of the probe laser $\Omega_{12}$ and known as EIT. For path shown with the red color, the control field $\Omega_{34}$ recovers the absorption against the EIT and known as EITA\cite{PAN13}. Similarly, the control fields $\Omega_{45}$ and $\Omega_{56}$ causes EITAT and EITATA \cite{PAN13} expressed by \textcolor{red}{EITATA1} in Eq. [1]. The other path shown with green color will also cause EITATA by sequence of the control fields $\Omega_{36}$ $\Omega_{56}$, $\Omega_{45}$ and $\Omega_{34}$ which is expressed by \textcolor{green}{EITATA2}. The other term in this expression is the interference term between the two sub-systems and is phase($\phi$) dependent.

\begin{figure}
   \begin{center}
      \includegraphics[width =\linewidth]{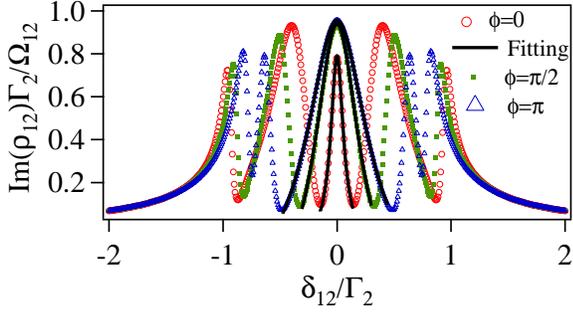}
      \caption{(Color online). Normalized absorption (Im($\rho_{12}$)$\Gamma_2/\Omega_{12}$) vs $\delta_{12}/\Gamma_2$ of the probe laser with $\delta_{23}=\delta_{34}=\delta_{45}=\delta_{56}=\delta_{36}=0$, $|\Omega_{23}|=|\Omega_{34}|=|\Omega_{45}|=|\Omega_{56}|=\Gamma_2$ and $|\Omega_{36}|=0.5\Gamma_2$.}
      \label{Abs_various_phase}
   \end{center}
\end{figure}

First we investigate the normalized absorption  (Im($\rho_{12}$)$\Gamma_2/\Omega_{12}$) vs probe detuning ($\delta_{12}$) for three different phases, $\phi=0,\pi/2,\pi$ as shown in Fig. \ref{Abs_various_phase}. For the central absorption peak i.e. at $\delta_{12}=0$, only the linewidth depends upon the phase but not the position, while both the position and the linewidth depends upon the phase($\phi$) for the other four absorption peaks. At high Rabi frequencies (much greater than the absorption peaks linewidths) of the control lasers and fields, the linewidth can be explained using dressed state picture. For general control fields detunings and Rabi frequencies, the position of the absorption peaks will be complicated. However, the expression becomes simpler for zero detuning of control fields where $|\Omega_{23}|=|\Omega_{34}|=|\Omega_{45}|=|\Omega_{56}|=\Omega$ but different value of $\Omega_{36}$ whose phase is to be determined. In this condition the positions of the absorption peaks are 
\tiny{}$-\sqrt{4\Omega^2+|\Omega_{36}|^2+\sqrt{(2\Omega^2+|\Omega_{36}|^2)^2+8\Omega^3|\Omega_{36}|cos\phi}}$,\\
$-\sqrt{4\Omega^2+|\Omega_{36}|^2-\sqrt{(2\Omega^2+|\Omega_{36}|^2)^2+8\Omega^3|\Omega_{36}|cos\phi}}$, 0,\\
$\sqrt{4\Omega^2+|\Omega_{36}|^2-\sqrt{(2\Omega^2+|\Omega_{36}|^2)^2+8\Omega^3|\Omega_{36}|cos\phi}}$,\\
\normalsize{} and \tiny{} $\sqrt{4\Omega^2+|\Omega_{36}|^2+\sqrt{(2\Omega^2+|\Omega_{36}|^2)^2+8\Omega^3|\Omega_{36}|cos\phi}}$.       
\normalsize{}The central peak is a superposition of the bare atomic states and is expressed as \tiny{}$[(1-e^{i\phi}\frac{\Omega_{36}}{\Omega})\ket{2}-\ket{4}+\ket{6}]$/$[(1+\frac{\Omega^2_{36}}{\Omega^2}-2\frac{\Omega_{36}}{\Omega}cos\phi)+2]^{1/2}$.\normalsize{} It's linewidth is given by \tiny{}$[(1+\frac{\Omega^2_{36}}{\Omega^2}-2\frac{\Omega_{36}}{\Omega}cos\phi)\Gamma_2+\Gamma_4+\Gamma_6]$/$[(1+\frac{\Omega^2_{36}}{\Omega^2}-2\frac{\Omega_{36}}{\Omega}cos\phi)+2]$ \normalsize{}which is phase dependent. In order to crosscheck the expression for the linewidth, we fit (shown with black solid line) the central peak of the normalized absorption obtained by Eq. [1] with Lorentzian profile to find the linewidth for three different phases as shown in Fig. \ref{Abs_various_phase}. The fitted linewidths for $\phi=0$, $\phi=\pi/2$ and $\phi=\pi$ are $0.13\Gamma_2$, $0.47\Gamma_2$ and $0.64\Gamma_2$ respectively, while the calculated  linewidths are $0.13\Gamma_2$, $0.39\Gamma_2$ and $0.54\Gamma_2$ respectively.
There is a small mismatch between the fitted and the calculated linewidths for $\phi=\pi/2$ and $\phi=\pi$. As we see in Fig. \ref{Abs_various_phase}, the central absorption peak is broadened for $\phi=\pi/2$ and $\phi=\pi$ and the interference between peaks starts playing a role in the modification of the linewidth.      

\begin{figure}
   \begin{center}
      \includegraphics[width =\linewidth]{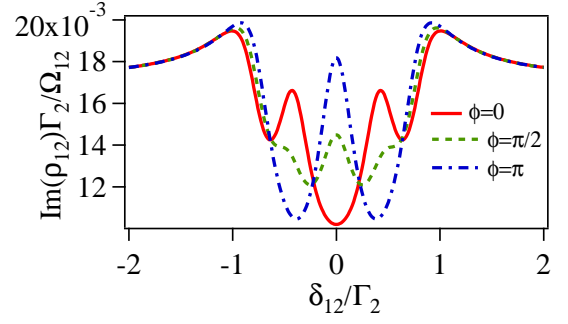}
      \caption{(Color online). Normalized absorption of the probe laser with thermal averaging (Im($\rho^{\textrm{Thermal}}_{12}$)$\Gamma_2/\Omega_{12}$) vs $\delta_{12}/\Gamma_2$ with $\delta_{23}=\delta_{34}=\delta_{45}=\delta_{56}=\delta_{36}=0$, $|\Omega_{23}|=|\Omega_{34}|=|\Omega_{45}|=|\Omega_{56}|=\Gamma_2$ and $|\Omega_{36}|=0.5\Gamma_2$.}
      \label{Abs_various_phase_Dopp}
   \end{center}
\end{figure}

Lets consider the effect of the temperature as lineshape of EIT is significantly changed by the thermal averaging \cite{KPW05,MJA07,KAV08,LNM15,IFN09,PKP16}.  
The thermal averaging of $\rho_{12}$ is done numerically for the room temperature ($T=$300~K) for the counter-propagating configuration of the probe ($\Omega_{12}$) and the control laser($\Omega_{23}$) with wave-vectors $k_{780}$ and $k_{480}$ respectively  by replacing $\delta_{12}$ with $\delta_{12}+k_{780}v$ and $\delta_{23}$ with $\delta_{23}-k_{480}v$ for moving atoms with velocity $v$ in Eq. [1] while the Doppler shift for the MW or RF field is ignored. Further the $\rho_{12}$ is weighted by the Maxwell Boltzman velocity distribution function and integrated over the velocity as $\rho^{\textrm{Thermal}}_{12}=\sqrt{\frac{m}{2\pi k_B T}}\int^{500\textrm{m/s}}_{-500\textrm{m/s}}\rho_{12}(v)e^{-\frac{mv^2}{2k_B T}}dv$, where $k_B$ is Boltzman constant and $m$ is atomic mass of Rb. The integration is done over velocity range which is three times of $\sqrt{\frac{k_BT}{m}}$. The Doppler averaging changes the absorption profile significantly as shown in Fig. \ref{Abs_various_phase_Dopp}. One of the interesting modification is the phase dependency of the probe laser absorption at the zero detunings of the probe, control laser and MW fields. The probe laser absorption is minimum for $\phi=0$ and maximum for $\phi=\pi$ as shown with red and blue curve respectively in Fig. \ref{Abs_various_phase_Dopp}. This modification is due to mismatch of Doppler shift for probe at 780~nm and the control at 480~nm for moving atom. The modification of lineshape of EIT due to Doppler mismatch between probe and control has been seen previously \cite{KPW05,MJA07,KAV08,LNM15}. 
However, please note that without thermal averaging at zero detunings of the probe, control laser and MW fields, probe laser absorption has no significant difference between $\phi=\pi/2$ and $\pi$. In fact if we take $\gamma^{dec}=0$ there will be no phase dependency at all on the probe absorption in this particular condition.

\begin{figure}
   \begin{center}
      \includegraphics[width =\linewidth]{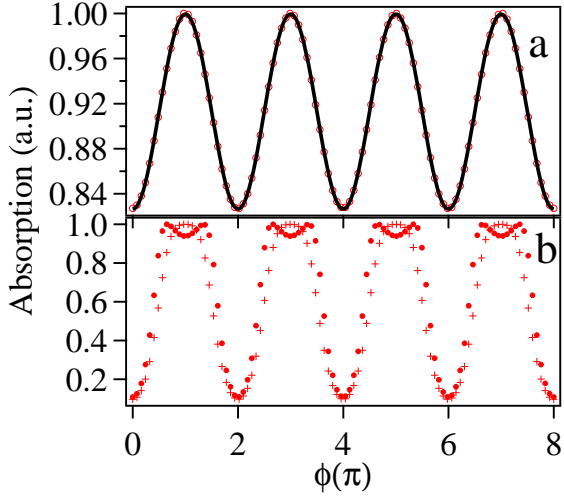}
     \caption{(Color online). Absorption of the probe laser after thermal averaging in arbitrary scale obtained as (Im($\rho^{\textrm{Thermal}}_{12}$))/max(Im($\rho^{\textrm{Thermal}}_{12}$)) with $\delta_{12}=\delta_{23}=\delta_{34}=\delta_{45}=\delta_{56}=\delta_{36}=0$ and (a) $|\Omega_{36}|=0.1\Gamma_2$, $|
\Omega_{23}|=2\Gamma_2$, $|\Omega_{34}|=1.5\Gamma_2$, and $|\Omega_{45}|=|\Omega_{56}|=4\Gamma_2$. (b) crossed points $|\Omega_{36}|=2.5\Gamma_2$, $|\Omega_{23}|=3\Gamma_2$, $|\Omega_{34}|=2\Gamma_2$, and $|\Omega_{45}|=|\Omega_{56}|=4\Gamma_2$, solid circled points $|\Omega_{36}|=2.5\Gamma_2$, $|\Omega_{23}|=3\Gamma_2$, $|\Omega_{34}|=3\Gamma_2$, and $|\Omega_{45}|=|\Omega_{56}|=4\Gamma_2$.}
      \label{Abs_vs_Phase}
   \end{center}
\end{figure}

Further, we explore the probe absorption at room temperature vs the phase $\phi$ with all the detunings to be zero. 
%From the plot shown with red open circle in Fig. \ref{Abs_vs_Phase}a we see more than 15$\%$ change in the probe absorption for the change of the phase from $0$ to $\pi$ for the chosen combinations of the control Rabi frequencies. The numerical data points (red circled) are fitted by a function A+Bsin(f$\phi$+$\theta$), where A, B, f and $\theta$ are kept as free parameters that yields f=1 and the fitting is shown with black curve in Fig. \ref{Abs_vs_Phase}a. For high values of the Rabi frequencies of the control laser and fields, there is increase in the contrast of the absorption for the $\phi=0$ and $\phi=\pi$ but there is a deviation form sinusoidal behavior as shown with red solid points and fitted with  also a distortion from the sinusoidal behavior as shown in Fig. \ref{Abs_vs_Phase}b with red solid circled points and cross points.   
From the plot shown with red open circle in Fig. \ref{Abs_vs_Phase}a we observe more than 15$\%$ change in the probe absorption for the change of the phase from $0$ to $\pi$ for the chosen combinations of the control Rabi frequencies. In particular, we have chosen low value of $|\Omega_{36}|=0.1\Gamma_2$ and the optimized parameters value are $|\Omega_{23}|=2\Gamma_2$, $|\Omega_{34}|=1.5\Gamma_2$, and $|\Omega_{45}|=|\Omega_{56}|=4\Gamma_2$ for this plot.  The numerical data points (red open circle) are fitted by a function A+Bsin(f$\phi$+$\theta$), where A, B, f and $\theta$ are kept as free parameters that yields f=1 and the fitting is shown with black curve in Fig. \ref{Abs_vs_Phase}a. Now, choosing a high value of  $|\Omega_{36}|=2.5\Gamma_2$ and keeping the other parameters unchanged, we observe more than 80$\%$ change in the probe absorption for the change of the phase from $0$ to $\pi$, but there is a deviation from sinusoidal behavior. This deviation is shown by crossed red points and is compared with the fitted black curve as shown in Fig. \ref{Abs_vs_Phase}b. On increasing the value of $|\Omega_{34}|$ to $3\Gamma_2$ and keeping the other parameters unchanged, there is a splitting of the absorption at $\phi=\pi$ as shown by the solid circled points in this figure.

\begin{figure}
   \begin{center}
      \includegraphics[width =\linewidth]{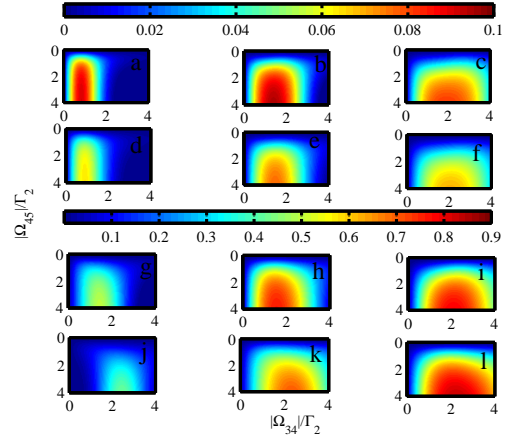}
      \caption{(Color online). Color map for $\phi_{\textrm{contrast}}$ vs $|\Omega_{34}|/\Gamma_2$ and $|\Omega_{45}|/\Gamma_2=|\Omega_{56}|/\Gamma_2$, for $\delta_{12}=\delta_{23}=\delta_{34}=\delta_{45}=\delta_{56}=\delta_{36}=0$, (a)-(c) $|\Omega_{36}|=0.1\Gamma_2$, $|\Omega_{23}|=\Gamma_2$, $2\Gamma_2$ and $3\Gamma_2$ with $\gamma^{dec}=2\pi\times$100kHz respectively. (d)-(f) $|\Omega_{36}|=0.1\Gamma_2$, $|\Omega_{23}|=\Gamma_2$, $2\Gamma_2$ and $3\Gamma_2$ with $\gamma^{dec}=2\pi\times$500kHz respectively. (g)-(i) $|\Omega_{36}|=1.5\Gamma_2$, $|\Omega_{23}|=\Gamma_2$, $2\Gamma_2$ and $3\Gamma_2$ with $\gamma^{dec}=2\pi\times$100kHz respectively, (j)-(l) $|\Omega_{36}|=2.5\Gamma_2$, $|\Omega_{23}|=\Gamma_2$, $2\Gamma_2$ and $3\Gamma_2$ with $\gamma^{dec}=2\pi\times$100kHz respectively. Note that the top most color bar corresponds to Fig. \ref{Abs_Contrast} (a)-(f) while the other one corresponds to  Fig. \ref{Abs_Contrast} (g)-(l).}
      \label{Abs_Contrast}
   \end{center}
\end{figure}

We further define a quantity called phase contrast as $\phi_{\textrm{contrast}}$=Im[$\rho^{\textrm{Thermal}}_{12}(\phi=0)-\rho^{\textrm{Thermal}}_{12}(\phi=\pi)$]/Im[$\rho^{\textrm{Thermal}}_{12}(\phi=0)+\rho^{\textrm{Thermal}}_{12}(\phi=\pi)$], which is a measure of the phase sensitivity. We optimize $\phi_{\textrm{contrast}}$ using various parameters. In Eq. [1] the expression for $\rho_{12}$ is symmetric in $|\Omega_{45}|$ and $|\Omega_{56}|$ so we keep $|\Omega_{45}|=|\Omega_{56}|$. For low value of the $|\Omega_{36}|$ ($=0.1\Gamma_2$) as shown in Fig. \ref{Abs_Contrast} (a)-(f),  the optimum value of $\phi_{\textrm{contrast}}$ is 0.096 which corresponds to $|\Omega_{23}|\approx2\Gamma_2$ and $|\Omega_{34}|\approx1.5\Gamma_2$ (Fig. \ref{Abs_Contrast}(b)) for the $\gamma^{dec}=2\pi\times100$ kHz. In order to see the effect of the $\gamma^{dec}$ for same low value of $|\Omega_{36}|$, we increase the value of former to $2\pi\times$500 kHz and we observe that the optimum value of $\phi_{\textrm{contrast}}$ is reduced to 0.081 (Fig. \ref{Abs_Contrast}(e)). For high value of the $|\Omega_{36}|$ ($=1.5\Gamma_2$) the optimum value of  $\phi_{\textrm{contrast}}$ is 0.79 which corresponds to the higher value of $|\Omega_{23}|\approx3\Gamma_2$ and $|\Omega_{34}|\approx2.1\Gamma_2$ (Fig. \ref{Abs_Contrast}(i)). For further higher value of the $|\Omega_{36}|$ ($=2.5\Gamma_2$) the optimum value of  $\phi_{\textrm{contrast}}$ is 0.83 which corresponds to the value of $|\Omega_{23}|\approx3\Gamma_2$ and $|\Omega_{34}|\approx2.3\Gamma_2$ (Fig. \ref{Abs_Contrast}(l)).

\begin{figure}
   \begin{center}
      \includegraphics[width =\linewidth]{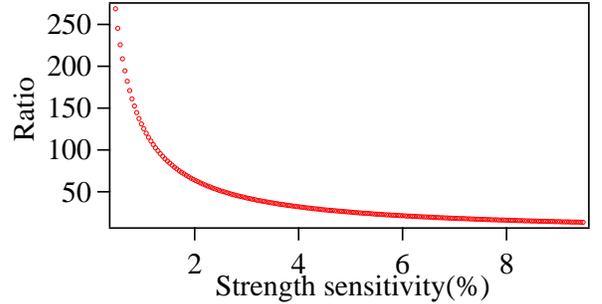}
      \caption{(Color online). Ratio vs sensitivity of the six level system for the optimum parameters.}
      \label{Enhancement_factor}
   \end{center}
\end{figure}

Now we compare the strength sensitivity for the MW field, between the previously studied four-level system and the six-level loopy ladder system. We define the $\%$ strength sensitivity as \small{}\big[$\rho^{\textrm{Thermal}}_{12}(|\Omega_{36}|\neq0$)-$\rho^{\textrm{Thermal}}_{12}(|\Omega_{36}|=0$)\big]\big/\big[$\rho^{\textrm{Thermal}}_{12}(|\Omega_{36}|\neq0$)+$\rho^{\textrm{Thermal}}_{12}(|\Omega_{36}|=0$)\big]$\times$100.
\normalsize{}
For low value of $|\Omega_{36}|$ ranging 0.05-0.1$\Gamma_2$  with optimum input parameters ($|\Omega_{45}|=|\Omega_{56}|=4\Gamma$,$|\Omega_{23}|=2\Gamma$, and $|\Omega_{34}|=1.5\Gamma$) the ratio of the percentage sensitivity between six-level and the four-level system is in the range of 10-250  as shown in Fig. \ref{Enhancement_factor}.

In conclusion we theoretically study a loopy ladder system using Rydberg states for the phase sensitive MW or RF electrometry. This is based upon the interference between the two sub-systems of EITATA.  In counter-propagating configuration of the probe and control laser there is a change of the lineshape of the probe absorption due to Doppler averaging. This makes detection of the phase possible at the zero detunings of the lasers and the fields.
This opens up a great possibility to characterize the MW or RF electric fields completely including the propagation direction and the wavefront. This system also provides orders of magnitude more sensitivity to the field strength as compared to previously the studied four level system. This work will be quite useful for MW and RF engineering hence in the communications and Radar technologies such as synthetic aperture radar interferometry.     

K.P. would like to acknowledge the discussion with David Wilkowski at CQT NTU and Sambit Bikas Pal at CQT NUS for this work.

\bibliography{eitrefsall}

\end{document}